# Realization scheme for visual cryptography with computer-generated holograms

Tao Yu,[1] Jinge Ma,[1] Guilin Li,[1] Dongyu Yang,[1] Rui Ma,[1] and Yishi Shi[1][2]

[1] School of Optoelectronics, University of Chinese Academy of Sciences, Beijing, 100049, People's Republic of China

[2] Academy of Opto-Electronics, Chinese Academy of Sciences, Beijing, 100049, People's Republic of China

**Abstract**

We propose to realize visual cryptography in an indirect way with the help of computer-generated hologram. At present, the recovery method of visual cryptography is mainly superimposed on transparent film or superimposed by computer equipment, which greatly limits the application range of visual cryptography. In this paper, the shares of the visual cryptography were encoded with computer-generated hologram, and the shares is reproduced by optical means, and then superimposed and decrypted. This method can expand the application range of visual cryptography and further increase the security of visual cryptography.

## 1. Introduction

Visual cryptography was proposed by Naor and Shamir at the 1994 European Conference on Cryptography[1]. Based on the principle of secret sharing of secrets, it combines secret sharing with digital images to form a new research hotspot. The secret sharing algorithm encodes the secret image into several images called Shares, and the black and white pixels in the shares are randomly distributed, from which no information about the secret image is obtained. The secret recovery algorithm is very simple, just print a certain number of shared copies to the transparencies and overlay them, and the human visual system can directly recognize the secret information.

Although visual cryptography have the secret security features of secret sharing, the hidden communication characteristics of shares, and the simplicity of superimposed visibility, many advantages and characteristics[2]. However, the current methods for recovering visual cryptography are mainly superimposed on transparent film or superimposed using computer equipment, which greatly limits the application range of visual cryptography. In response to this problem, other staff in our lab have done some research[3]~[6]. In this paper, the shares of the visual cryptography were encoded with computer-generated holograms (CGHs). In addition to using computer decryption, the shares can be reproduced optically and decrypted directly in space.

## 2. Proposed scheme

As shown in Fig. 1, the proposed scheme consists of the follow steps:

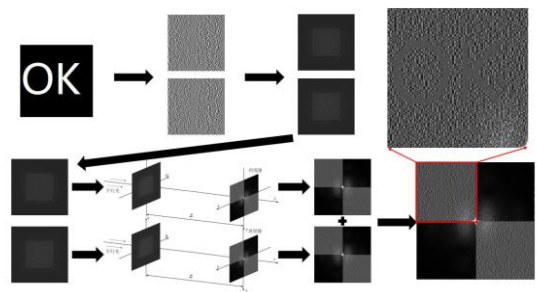

Fig. 1. Encryption and decryption process

Step 1: Generate shares. Taking the Shamir visual cryptography scheme as an example, the secret information is a combination of a series of black and white pixels, each of which is processed separately. Each original pixel is divided into *n* shares, each of which contains *m* sub-pixels that are very similar to each other, so that the human visual system averages their unique black or white contributions.

Step 2: Make computer-generated holograms. The CGH are created by taking the shares generated in the first step as objects.

Step 3: Reproduction and decryption. The CGH are reproduced by optical means to obtain shares, and superpose the shares, then the secret information is obtained.

## 3. Experiment

Burch coding CGHs[7] with Shamir visual cryptography scheme is used in our simulation experiments. The secret information is shown in Fig. 2(a). In encryption, two Burch coding CGH images are generated, as shown in Fig. 2(b) and 2(c).In decryption, two coding CGH images need to reconstruct. Then, reconstruction images are superimposed and the simulation result is shown in Fig. 2(c). As shown in Fig. 2(d), the decryption image from a part of Fig. 2(c), because of conjugate images of CGH.

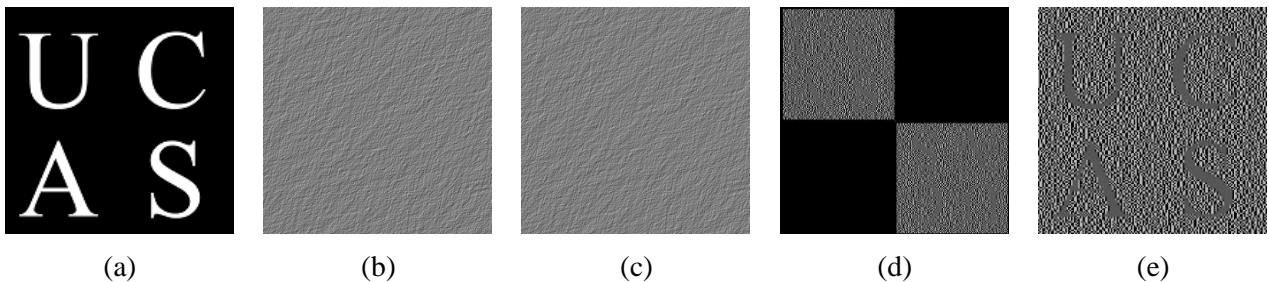

(a)      (b)      (c)      (d)      (e)

Fig. 2. (a) secret information, (b)(c) encryption images, (d) reconstruction decryption image, (e) is part of(d)

## 4. Conclusion

A new visual cryptography based on CGH is proposed. In this way, the secret information are able to encryption and decryption. And visual cryptography application field is extended, which is not limited in computer or film. Meanwhile, the security of visual cryptography is also increased by CGH coding.

■ Please send us this submission form with your manuscript.

# IWH2018 Paper Submission Form

| No | Title | Family Name | Given Name | Affiliation | Department | Address | Country | e-mail |
|---|---|---|---|---|---|---|---|---|
| | Mr. | Yu | Tao | School of Optoelectronics | University of Chinese Academy of Sciences | Beijing, 100049 | People's Republic of China | yutao4291@163.com |

| Presentation Title | Abstract (up to 35 Words) |
|---|---|
| Realization scheme for visual cryptography with computer-generated hologram | We propose to realize visual cryptography in an indirect way with the help of computer-generated hologram. At present, the recovery method of visual cryptography is mainly superimposed on transparent film or superimposed by computer equipment, which greatly limits the application range of visual cryptography. In this paper, the shares of the visual cryptography were encoded with computer-generated hologram, and the shares is reproduced by optical means, and then superimposed and decrypted. This method can expand the application range of visual cryptography and further increase the security of visual cryptography. |